\documentclass[]{spie}  %>>> use for US letter paper
\usepackage{amsmath,amsfonts,amssymb}
\usepackage{booktabs}
\usepackage{graphicx}
\usepackage[colorlinks=true, allcolors=blue]{hyperref}
\usepackage[T1]{fontenc}
\usepackage{enumitem}

\usepackage{wrapfig}

\usepackage{array}
\usepackage{makecell}

\title{CCAT: The Prime-Cam Instrument for the Fred Young Submillimeter Telescope -- Overview and Status}

\author[a,b]{Eve M. Vavagiakis}
\author[b]{Yuhan Wang}
\author[b]{Lawrence T. Lin}
\author[c]{Manuel Aravena}
\author[d]{Jason\,E.\,Austermann}
\author[e]{Frank Bertoldi}
\author[f]{James Burgoyne}
\author[b]{Victoria Butler}
\author[g]{Scott Chapman}
\author[h]{Steve\,K.\,Choi}
\author[i]{Dongwoo Chung}
\author[b]{Abigail Crites}
\author[e]{Ankur Dev}
\author[b]{Cody J. Duell}
\author[j]{Michel Fich} 
\author[k]{Laura\,Fissel}
\author[i]{Rodrigo\,G.\,Freundt}
\author[l]{Eliza Gazda}
\author[f]{Anthony I. Huber}
\author[m,n]{Doug Johnstone}
\author[b]{Ben\,Keller}
\author[h]{Paul\,Malachuk}
\author[b]{Alicia\,Middleton}
\author[a]{Jenna Moore}
\author[b]{Michael D. Niemack}
\author[o]{Thomas\,Nikola}
\author[p,q]{Yoko\,Okada}
\author[b]{Darshan A. Patel}
\author[a]{Tilak M. Patel}
\author[p,q]{Dominik A. Riechers}
\author[f]{Douglas\,Scott}
\author[i]{Gordon\,J.\,Stacey}
\author[b]{Benjamin\,J.\,Vaughan}
\author[b]{Samantha Walker}
\author[f]{Ruixuan\,(Matt)\,Xie}
\author[b]{the\,CCAT\,Collaboration}
\affil[a]{Department of Physics, Duke University, Durham, NC 27710, USA}
\affil[b]{Department of Physics, Cornell University, Ithaca, NY 14853, USA}
\affil[c]{Instituto de Estudios Astrofísicos, Facultad de Ingenieria y Ciencias, Universidad Diego Portales, Av. Ejercito 441, Santiago, Chile}
\affil[d]{National Institute of Standards and Technology, Quantum Sensors Division, Boulder, Colorado, USA}
\affil[e]{Argelander-Institut f\"ur Astronomie, Universit\"at Bonn, Auf dem H\"ugel 71, 53121 Bonn, Germany}
\affil[f]{Department  of Physics and Astronomy, University of British Columbia, Vancouver, Canada}
\affil[g]{Department of Physics and Atmospheric Science, Dalhousie University, Halifax, BC, Canada}
\affil[h]{Center for Experimental Cosmology and Instrumentation, Department of Physics and Astronomy, University of California, Riverside, CA 92521, USA}
\affil[i]{Department of Astronomy, Cornell University, Ithaca, NY 14853, USA}
\affil[j]{Department of Physics and Astronomy, University of Waterloo, Waterloo, Ontario, CA, N2L 3G1}
\affil[k]{Department of Physics, Engineering Physics and Astronomy, Queen’s University,  Kingston, ON, Canada}
\affil[l]{Department of Physics and Astronomy, UC Riverside, Riverside, CA, USA 92521}
\affil[m]{NRC Herzberg Astronomy and Astrophysics, 5071 West Saanich Road, Victoria, BC, V9E 2E7, Canada}
\affil[n]{Department of Physics and Astronomy, University of Victoria, Victoria, BC, V8P 5C2, Canada}
\affil[o]{Cornell Center for Astrophysics and Planetary Sciences, Cornell University, Ithaca, NY, USA 14853}
\affil[p]{Institute for Astrophysics, University of Cologne, Zülpicher Straße 77, 50937 Cologne, Germany}
\affil[q]{Cluster of Excellence ``Our Dynamic Universe'' (DYNAVERSE)}

\authorinfo{Further author information: Send correspondence to E.M.V.: E-mail: eve.vavagiakis@duke.edu}

\begin{document} 
\maketitle

\newpage
\begin{abstract}
Prime-Cam is a first-generation science instrument for the CCAT Observatory’s six-meter aperture Fred Young Submillimeter Telescope (FYST), under construction at an elevation of 5600\,m on Cerro Chajnantor in Chile’s Atacama Desert. Prime-Cam will deliver over ten times greater mapping speed at submillimeter wavelengths than current facilities for unprecedented broadband and spectroscopic measurements in windows between 1.4 -- 0.3\,mm (220 -- 850\,GHz). When fully populated, Prime-Cam will field over 100,000 kinetic inductance detectors across seven independently optimized instrument modules. With Prime-Cam, the CCAT Collaboration will address a suite of science goals, from Big Bang cosmology, to galaxy evolution and star formation over cosmic time. Prime-Cam is scheduled for integration in FYST in late 2026, followed by a year of early science observations with the 280 and 350 GHz instrument modules. We discuss the design and in-lab testing of the 1.8-m diameter Prime-Cam receiver and 280 GHz instrument module, and give an update on deployment status and early science plans.
\end{abstract}

\keywords{Cryogenics, Cryostat design, Superconducting detectors, Kinetic Inductance Detectors, Cosmic Microwave Background, Millimeter and Submillimeter astrophysics, Cosmology}

\section{INTRODUCTION}
\label{sec:intro}  % \label{} allows reference to this section

\begin{figure}[ht!]
    \centering
    \includegraphics[width=\linewidth]{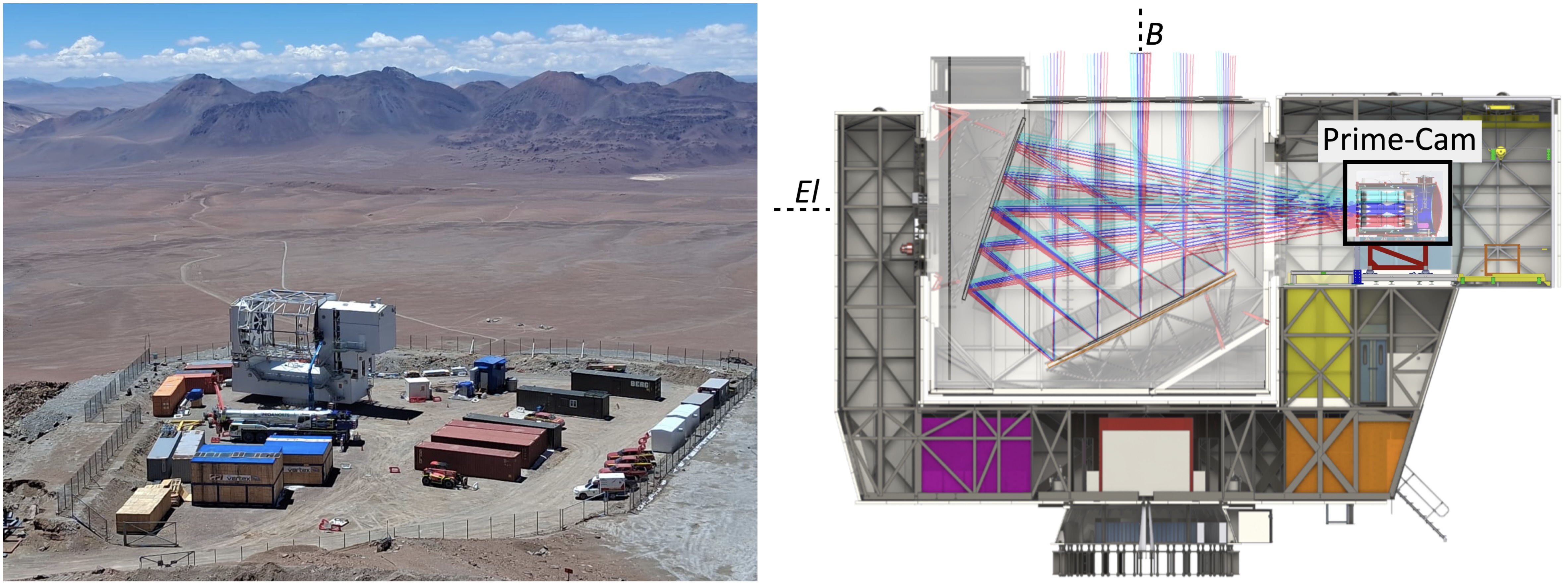}
    \caption{{\it Left:} The Fred Young Submillimeter Telescope under construction at 5600\,m on Cerro Chajnantor. {\it Right:} A cross-section of the FYST CAD model. Prime-Cam is in Instrument Space 1, mounted on the ``raft'', which interfaces with the focusing translation stage and instrument space floor. A ray trace of light from the sky reflecting off the 6-m primary and secondary mirrors in the telescope elevation housing is shown. The ray trace enters three Prime-Cam instrument modules and is focused onto the kinetic inductance detector (KID) arrays. When observing, the elevation housing rotates around the elevation axis (dashed black line ``{\it El\/}'') and the entire telescope around the support cone or boresight axis, ``{\it B\/}''.}
    \label{fig:fyst}
\end{figure}

The CCAT Collaboration, an international partnership of institutions led by Cornell University, is opening submillimeter windows to an unprecedented wide-field survey in the Chilean Atacama Desert with Prime-Cam on the 6-m aperture Fred Young Submillimeter Telescope (FYST, Fig.~\ref{fig:fyst}), designed and built by CPI Vertex Antennentechnik GmbH.\footnote{\url{www.ccatobservatory.org}}\footnote{\url{https://www.vertexant.com/en/}} FYST's wide diffraction-limited field of view (FoV, 2$^\circ$ diameter at 860\,GHz, 8$^\circ$ diameter at 100\,GHz) will enable faster mapping of the 220--850\,GHz sky than any current submillimeter facility \cite{Parshley_2018,Stacey2018,Niemack_2016,Parshley2018OpticalDesign}. FYST's design achieves low emissivity ($\sim 1\%$) and high mirror surface accuracy (half-wavefront error $\leq 10.7\,\mu$m) to take advantage of the 5600-m elevation Cerro Chajnantor site \cite{Parshley_2018}. This will enable a unique wide-field submillimeter survey from the Atacama, fully overlapping with the footprint of the Simons Observatory (SO) Large Aperture Telescope survey \cite{ASO2025,SO_Ade_2019,Zhu_2021} for exceptional cross-correlation opportunities \cite{ccat2023,Vavagiakis_2025}. 

Prime-Cam is a first-light camera designed for a broad suite of science goals. FYST will illuminate \textgreater 100,000 kinetic inductance detectors in Prime-Cam when fully populated to provide a detailed view into cosmic history \cite{Choi_2020,huber2024ccatprimecamopticsoverview,Vavagiakis_2018}. Prime-Cam accommodates up to seven instrument modules\cite{Zhu_2021}, each designed for a particular CCAT science goal, ranging from Big Bang cosmology to the physics of our own Milky Way. A set of wide-area and targeted surveys, beginning with one year of early science observations, will investigate \cite{ccat2023}: 

\vspace{-0.2cm}
\begin{itemize}[noitemsep]
    \item Polarized cosmic microwave background (CMB) foreground measurements to enable more precise constraints of primordial gravitational waves and light relic particles;
    \item Dark energy, large scale structure and galaxy evolution through the Sunyaev-Zeldovich (SZ) effects;
    \item Rayleigh scattering in the CMB;
    \item The formation and large-scale, three-dimensional evolution of the first star-forming galaxies during the Epoch of Reionization;
    \item The evolution of dusty star-forming galaxies since the epoch of galaxy assembly;
    \item Star formation, magnetic fields, turbulence, and dust grain properties in the Milky Way and nearby galaxies;
    \item The time-variable submillimeter sky, including monitoring the Galactic center and Galactic star-forming regions for energetic transients.
\end{itemize}
\vspace{-0.25cm}

\vspace{0.125cm}
The status and descriptions of the seven Prime-Cam instrument modules are listed in Table~\ref{tab:moduleoverview}. Prime-Cam has been shipped to the CCAT site for integration on FYST and a one-year early science campaign with the lab-characterized 280 GHz and 350 GHz broadband polarimetric kinetic inductance detector (KID) modules. This paper describes the cryomechanical design of Prime-Cam (Sec.~\ref{sec:cryo}) and gives a brief overview of the instrument modules (Sec.~\ref{sec:modules}) and detectors and readout (Sec.~\ref{sec:dets}). Results from in-lab testing of the Prime-Cam receiver and cryogenic tests of the 280 GHz module are presented in Sec.~\ref{sec:tests}. Early science plans with Prime-Cam on FYST are detailed in Sec.~\ref{sec:earlysci}, with a discussion of commissioning and calibration plans in Sec.~\ref{sec:commissioning}, scheduling of surveys in Sec.~\ref{sec:scheduling}, and plans for each science case in Sec.~\ref{sec:scigoals}. Data management plans are described in Sec.~\ref{sec:datahandling}.

\vspace{0.5cm}
\begin{table*}[h!]
\centering
\caption{Summary of Prime-Cam instrument module frequency coverage, detector count for three KID arrays, resolution, and timeline. The 280, 350, 410, and 850 GHz modules are broadband polarimetric, while the EoR-Spec module is spectroscopic and not polarization-sensitive.}
\vspace{0.125cm}
\begin{tabular}{cccccc}
\toprule
Module &
Frequencies &
Detectors &
Resolution &
Status & 
Obs.~Date \\
& [GHz] & &
[arcsec] & & \\
\midrule

280\,GHz &
280 Broadband &
10,332 &
45 &
Shipped &
2026 \\

350\,GHz &
350 Broadband &
10,448 &
35 &
Shipped &
2026 \\

EoR-Spec&
210--420 Spectroscopic&
6,528 &
30--57 &
Development &
2027 \\

850\,GHz&
850 Broadband &
38,000 &
14 &
Development &
2027 \\

410\,GHz&
410 Broadband &
21,000 &
30 &
Development &
2028 \\

Spec-on-Chip &
Spectroscopic &
TBD &
TBD &
Design &
2028 \\

TBD &
TBD &
TBD &
TBD &
Planning & 
TBD \\

\bottomrule
\end{tabular}
\vspace{0.125cm}
\label{tab:moduleoverview}
\end{table*}

\section{The Prime-Cam Receiver}

Prime-Cam is a 1.8-m diameter receiver (Sec.~\ref{sec:cryo}), designed after the SO Large Aperture Telescope Receiver \cite{Zhu_2021}, that will fill the central 4.9$^{\circ}$ of FYST's FoV with seven instrument modules (Sec.~\ref{sec:modules}, Fig.~\ref{fig:fyst}) \cite{huber2024ccatprimecamopticsoverview,Choi_2020,Vavagiakis_2018,ccat2023}. When fully populated, the modules will deploy over 100,000 KIDs (Sec.~\ref{sec:dets}). Prime-Cam has been shipped to the Cerro Chajnantor site with the 280 GHz and 350 GHz instrument modules. We present the laboratory validation of the receiver and 280 GHz module in Sec.~\ref{sec:tests}. The 350 GHz module testing is presented in Ref.~\citenum{Keller2026}.

\begin{figure}
    \centering
    \includegraphics[width=0.5\linewidth]{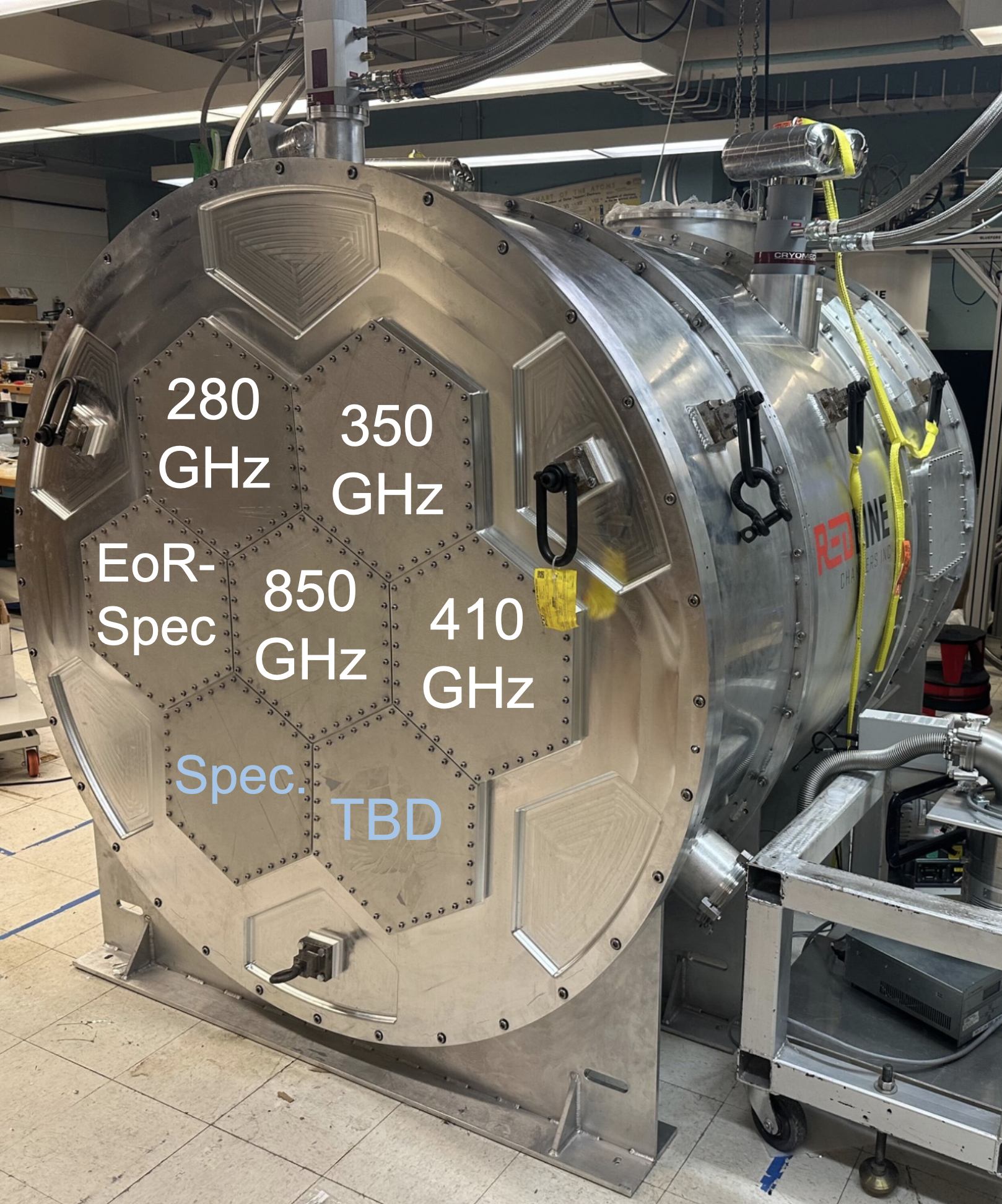}
    \caption{Prime-Cam in testing at Cornell University with vacuum window blank-offs and in a pulse tube (PT)-only configuration. The planned positions of instrument modules are labeled, with funded modules in development or shipped to the site labeled in white, and planned future modules in light blue.}
   \label{fig:modulelabelsprimecam}
\end{figure}

\subsection{Mechanical Design}
\label{sec:cryo}

\begin{figure}[t]
    \centering
    \includegraphics[width=0.75\linewidth]{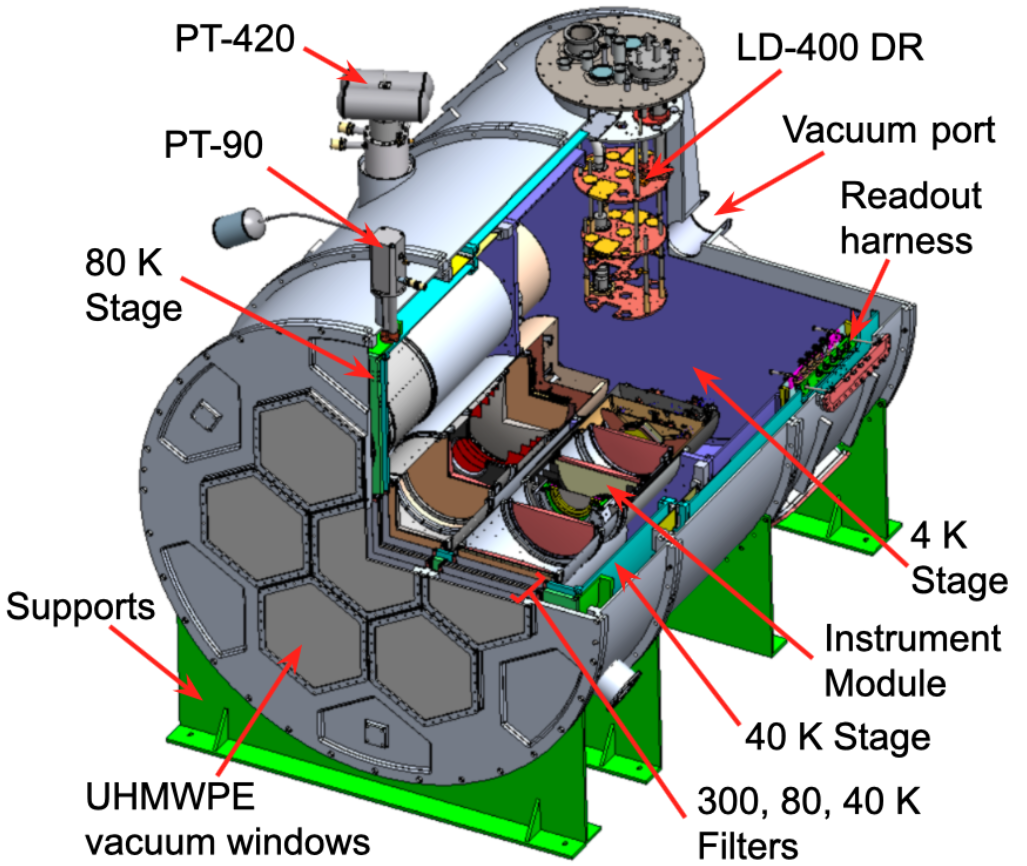}
    \vspace{0.125cm}
    \caption{Cutaway rendering of the Prime-Cam cryostat with instrument modules installed, showing optical elements and detector arrays. The 300 K stage is shown in grey, the 80 K optical filter stage in green, the 40 K stage in light blue, and the 4 K stage in purple. G10 tabs can be seen in yellow. The thermal connections between the 1 K and 100 mK DR plates and the instrument module cold fingers are not shown, nor is the dilution refrigerator turbo pump rack. The receiver supports (green) will couple the receiver to the raft (Fig.~\ref{fig:primecamonraft}) for installation in the telescope instrument space. Figure from Ref.~\citenum{Vavagiakis_2025}.}
    \label{fig:primecam_cutaway}
\end{figure}

The Prime-Cam cryostat (Fig.~\ref{fig:modulelabelsprimecam}) was manufactured by Redline Chambers\footnote{\url{https://www.redlinechambers.com}}, and includes aluminum 300 K, 80 K, 40 K and 4 K stages (Fig.~\ref{fig:primecam_cutaway}). All metal parts are Al 6061-T6 except for the 80 K front plate and 40 K front and back shells, which are made of Al 1100 for improved thermal conductivity. Prime-Cam's thermal stages and instrument module components are cooled by three pulse tube cryocoolers and a Bluefors\footnote{\url{https://bluefors.com}} LD400 dilution refrigerator (DR). Key elements of each stage are detailed below. 

\begin{figure}
    \centering
    \includegraphics[width=\linewidth]{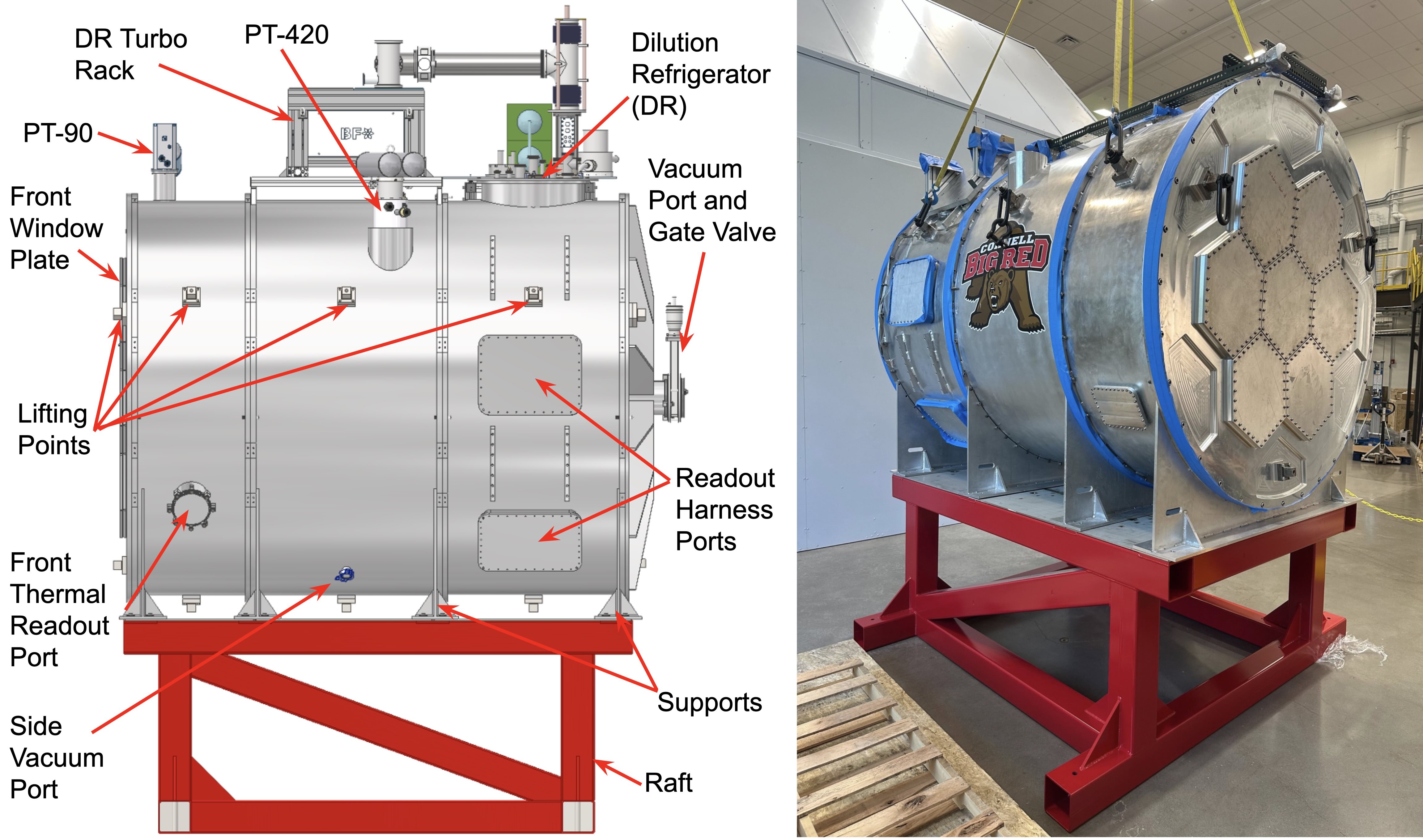}
    \caption{{\it Left:} Rendering of Prime-Cam on the raft. Key components of the external 300 K stage are labeled. {\it Right:} Final fit check of Prime-Cam on the raft before shipping both to Chile. The DR, DR turbo rack, and other components were removed for this fit check.}
    \vspace{0.125cm}
    \label{fig:primecamonraft}
\end{figure}

The 300 K stage is detailed in Fig.~\ref{fig:primecamonraft}. Prime-Cam interfaces to the telescope instrument space (Fig.~\ref{fig:fyst}) through four supports mounted to the 300 K shell flanges. The 300 K shell flanges are mounted to the ``raft", a stainless steel welded platform, which is mounted to the translation stage. Prime-Cam is focused relative to the telescope's mirrors in the instrument space by moving the translation stage along the floor of the instrument space, while Prime-Cam, its supports, and the raft remain fixed relative to this stage. Light from FYST enters Prime-Cam through anti-reflection-coated ultra-high molecular weight polyethylene windows. Directly behind each window on the 300 K front plate is a double-sided infrared-blocking filter (DSIR), which reduces optical loading on the colder stages \cite{Ade2006MetalMeshFilters}.

The 80 K stage is coupled to a PT-90 Cryomech\footnote{\url{https://bluefors.com/products/cryomech-products/}} cryocooler via Oxygen-Free High Thermal Conductivity (OFHC) Copper straps from Technology Applications, Inc.\ (TAI)\footnote{\url{https://www.techapps.com}}, and holds DSIR blocking filters as well as alumina wedge filters. Because the focal plane of the two-mirror crossed-Dragone telescope is curved, the rays entering modules other than Prime-Cam's central module are not parallel with the central axis of Prime-Cam \cite{Dicker2018ColdOpticalDesign}. The central alumina filter is thus flat, while the outer alumina filters are designed to act as prisms to refract incoming light into the instrument modules and enable a flat detector focal plane. The 80 K stage is mechanically supported and thermally isolated from the 300 K stage by twelve epoxied G10 fiberglass tabs. To reduce thermal loading, the 80 K stage is wrapped in 30 layers of multi-layer insulation (MLI).

The 40 K stage is comprised of front and back shells, a G10 mounting ring, a front plate holding one DSIR filter per module, and a back plate. The rear shell includes a 40 K readout harness plate interface. The G10 mounting ring is mounted to a series of G10 tabs to hold the 40 K stage off of the 300 K stage, and the 4 K stage off the 40 K stage. The 40 K shell is cooled by two PT-420 Cryomech cryocoolers via straps from TAI, and is wrapped with 30 layers of MLI. The dilution refrigerator is mounted through the top of the 40 K shell, and is not thermally coupled to the 40 K shell beyond MLI to close the optical path around the 40 K DR stage. 

\begin{figure}[p]
    \centering
    \includegraphics[width=0.975\linewidth]{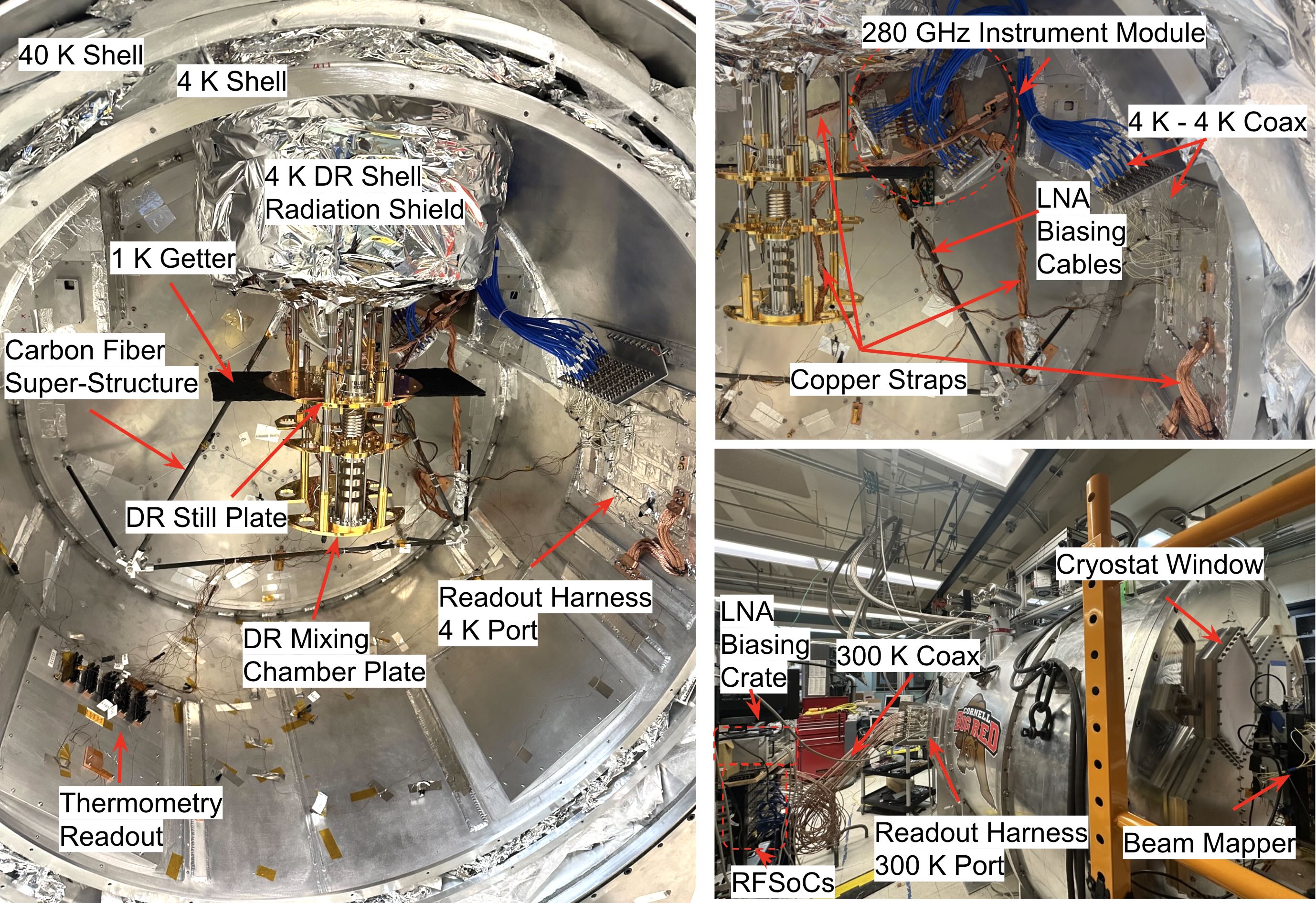}
    \caption{The 280 GHz module installed in Prime-Cam for testing. {\it Left}: With the DR installed, thermal straps connect the cold stages of the DR to the instrument module cold flanges. Carbon fiber supports and readout components connected for the 280 GHz module tests are visible within the 4 K radiation environment. {\it Right, top}: A clearer view of the 280 GHz module mounted to Prime-Cam's 4 K plate. {\it Right, bottom}: Beam mapping the 280 GHz module within Prime-Cam, with warm readout electronics noted.}
    \label{fig:280inprimecam}
%\end{figure}
\vspace{0.25cm}
%\begin{figure}
    \centering
    \includegraphics[width=0.85\linewidth]{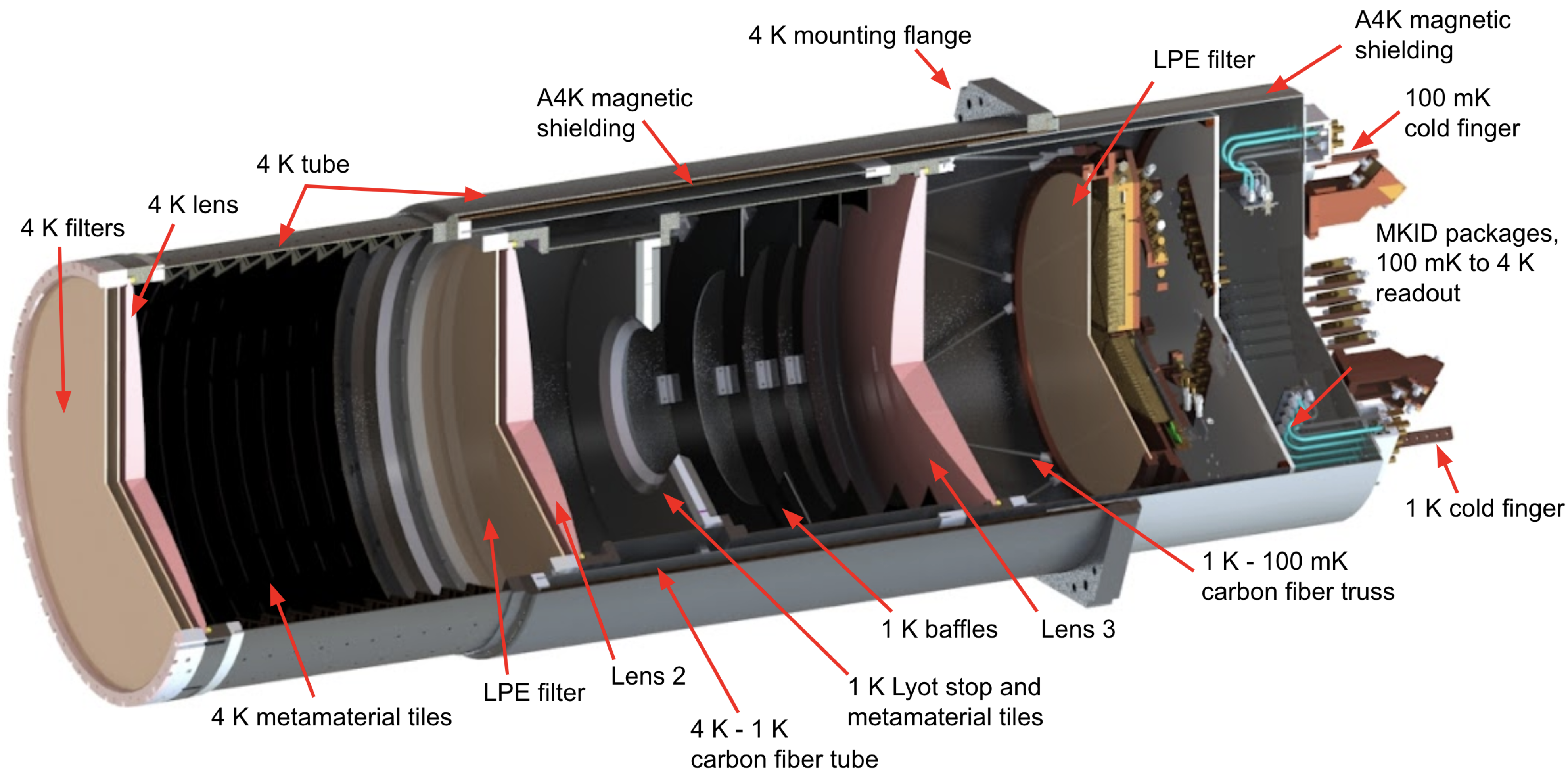}
    \caption{Rendered section view of the 280 GHz instrument module model from Ref.~\citenum{vavagiakis2022ccatprimedesignmodcamreceiver}. Optical elements at 4\,K, 1\,K and 100\,mK are labeled. The 1-K and 100-mK cold fingers, which thermally connect to the cold stages of the DR, are noted to the right, and can be seen inside Prime-Cam in Fig.~\ref{fig:280inprimecam}.}
    \vspace{0.125cm}
    \label{fig:instmod}
\end{figure}

The 4 K stage mechanically supports the instrument modules, which are inserted from the rear of the cryostat and cantilevered on the 4 K plate. The cold heads of two PT-420 Cryomech cryocoolers are heat sunk to the 4 K plate by TAI straps. The 4 K shell includes four readout harness plate interfaces, the inside of which can be seen in Fig.~\ref{fig:280inprimecam}. The 4 K stage of the dilution refrigerator is wrapped with MLI to close the optical path around the 4 K stage.

The 1 K and 100 mK instrument module stages (Sec.~\ref{sec:modules}) are cooled by the DR. At 1\,K, two carbon-loaded epoxy blackened absorbers are mounted to reduce the radiation temperature inside the 4 K shell environment and act as a getter for residual gas in the system. The cold plates of the DR are thermally coupled to the instrument module cold fingers through custom in-house fabricated OFHC copper straps supported by a carbon fiber truss (Fig.~\ref{fig:280inprimecam}).

\subsection{Instrument Modules}
\label{sec:modules}

Instrument modules being developed for Prime-Cam are listed in Table~\ref{tab:moduleoverview} and include the 280 \cite{lin2025ccatmodcamcryogenicperformance,vavagiakis2022ccatprimedesignmodcamreceiver,Duell_2020}, 350 \cite{Keller2026,huber2024ccatprimecamopticsoverview}, 410 \cite{Patel2026,Chapman2026}, and 850 GHz \cite{huber2022ccatprimeopticalcryogenicdesign,chapman2022ccatprime850ghzcamera} broadband polarimetric instrument modules and the Epoch of Reionization Spectrometer module (EoR-Spec) \cite{Freundt_2024,Nikola2023}, which is equipped with a Fabry-Perot interferometer. The Prime-Cam instrument modules were designed based on the SO Large Aperture Telescope Receiver (LATR) optics tube designs \cite{Zhu_2021}. Each instrument module has a 1.3$^{\circ}$ FoV, and contains optical elements and detectors from 4\,K to 100\,mK (Fig.~\ref{fig:instmod}). The optical designs vary between modules, but their common properties are described here.

The 4 K instrument module components include metamaterial anti-reflection coated silicon lenses and optical filters, as well as anti-reflection metamaterial tiles and magnetic shielding. The 1 K stage is supported off the 4 K stage by an epoxied carbon fiber shell, and holds lenses, filters, and optical elements such as blackened baffles. The 100 mK stage is supported from the 1 K stage by a carbon fiber truss, and holds the detector array packages (Sec.~\ref{sec:dets}). Gold-plated 100 mK and 1 K instrument module cold fingers couple to the cold stages of the DR, and readout electronics extend from the back of the modules, to read out signals through Prime-Cam's four readout harness ports. For more details on the 280 GHz instrument module design, see Ref.~\citenum{vavagiakis2022ccatprimedesignmodcamreceiver}.

\begin{figure}
    \centering
    \includegraphics[width=0.9\linewidth]{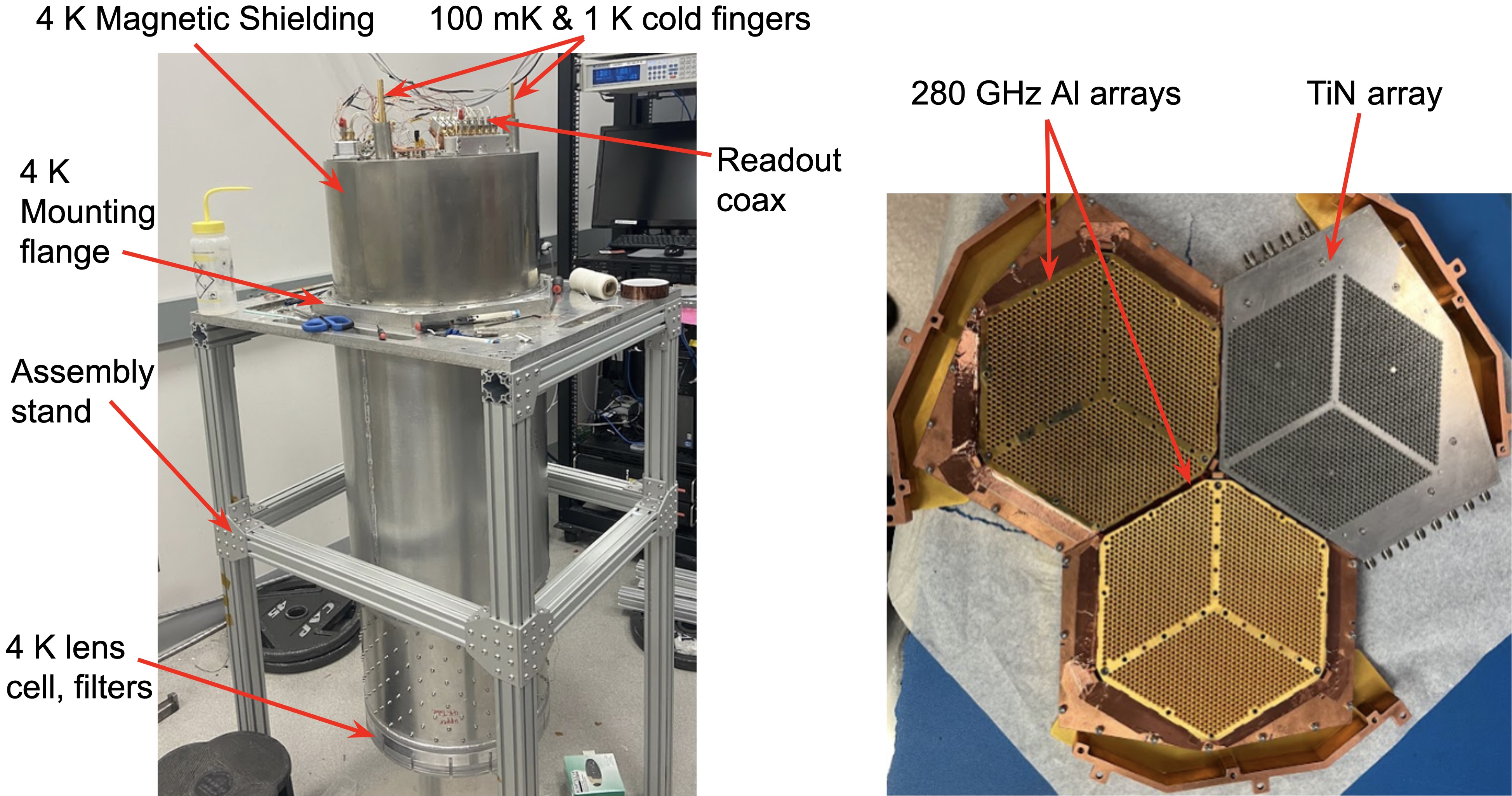}
    \vspace{0.125cm}
    \caption{{\it Left}: The 280 GHz module on the assembly stand. {\it Right}: The 280 GHz focal plane array, showing two gold-plated Si-platelet feedhorn assemblies for the Al arrays, and one Al-feedhorn assembly for the TiN array.}
    \vspace{0.125cm}
    \label{fig:280assembled}
\end{figure}

Modules undergo testing in the single-module-capacity Mod-Cam receiver before deployment in Prime-Cam \cite{lin2025ccatmodcamcryogenicperformance,keller2026ccatmodcamreadoutoverview,vavagiakis2022ccatprimedesignmodcamreceiver}. The 280 GHz instrument module (Fig.~\ref{fig:280assembled}) was also optically tested in Prime-Cam to validate the system before shipment (Sec.~\ref{sec:tests}).

\subsection{Detectors and Readout}
\label{sec:dets}

Significant developments in detector and readout technologies have paved the way towards fielding the largest number of KIDs in a single receiver. KIDs for Prime-Cam are being developed at the National Institute of Standards and Technology (NIST) Boulder for the 280 GHz (Fig.~\ref{fig:280assembled}) \cite{Vaskuri2025,lin2025ccatmodcamcryogenicperformance,Duell_2020,vavagiakis2022ccatprimedesignmodcamreceiver}, 350 GHz \cite{Keller2026}, 410 GHz \cite{Chapman2026}, 850 GHz \cite{chapman2022ccatprime850ghzcamera,chapman2022ccatprime850ghzcamera,huber2022ccatprimeopticalcryogenicdesign} and EoR-Spec modules \cite{Freundt_2024}. The number of KIDs per module across three detector arrays are listed in Table~\ref{tab:moduleoverview}. Detector arrays developed for Prime-Cam undergo extensive testing \cite{lin2025ccatmodcamcryogenicperformance,Vaskuri2025,huber2024ccatprimecamopticsoverview}, LED mapping \cite{Middleton2025}, capacitor trimming \cite{liu2017superconductingmicroresonatorarraysideal}, and characterization \cite{vaughan2026ccatmagneticsensitivitymeasurements,duell2024ccatcomparisons280ghz} efforts before deployment to FYST, including those presented here in Sec.~\ref{sec:tests}.

Over 1000 KIDs can be read out on a single radio frequency (RF) transmission line, and sophisticated warm readout electronics and software control systems have been developed for Prime-Cam. Custom readout harnesses were developed from the SO design \cite{moore2022developmentperformanceuniversalreadout} to read out detector timestream data. Each of Prime-Cam's readout harnesses is designed to accommodate 54 readout chains, with six chains per harness as spares \cite{keller2026ccatmodcamreadoutoverview,patel2025ccatreadout10000280} (Fig.~\ref{fig:280inprimecam}). Prime-Cam uses custom-packaged FPGA-based Xilinx ZCU111 radio frequency system on chip (RFSoC) systems. Each RFSoC can simultaneously read out four RF channels with up to 1,000 detectors spanning a 512 MHz bandwidth per channel using the current firmware \cite{SinclairSPIE2022}. The development of a two-octave 1.024 GHz KID readout with an
overlap-channel polyphase synthesis filter bank is also underway \cite{XieSPIE2026}. Custom readout software interfaces with these boards via dedicated Ethernet lines for tasks like tone comb driving, comb calibration and optimization, and detector timestream establishment \cite{burgoyne2024ccatfystprimecamreadout}.

\section{Laboratory Testing}
\label{sec:tests}

The Prime-Cam vacuum shells were delivered to Cornell University in February 2025 and subsequently integrated, modified, and tested in the laboratory. In February 2026, Prime-Cam was determined to have met the deployment requirements. By June 2026, the instrument had been disassembled, packaged, and shipped to the CCAT site together with the 280 and 350 GHz instrument modules in preparation for deployment. We present two highlights from Prime-Cam laboratory testing with the fully populated 280 GHz instrument module, including all three detector arrays and the complete optics set, as representative demonstrations that the instrument met the deployment criteria for cryogenic performance, readout readiness, and optical readiness.

\subsection{Cryogenic Testing}

As described above, Prime-Cam is cooled by one PT-90, two PT-420s, and one Bluefors LD400 DR. The cooldown process occurs in two stages. First, the receiver is cooled such that its coldest stage reaches approximately 4\,K, with the majority of the cooling power provided by the PT-420s and PT-90. After the 4 K, 1 K, and 100 mK stages of Prime-Cam stabilize, the DR mixture is condensed, cooling the DR mixing chamber and instrument module focal plane to below 100\,mK.

\begin{figure}
    \centering
    \includegraphics[width=0.9\linewidth]{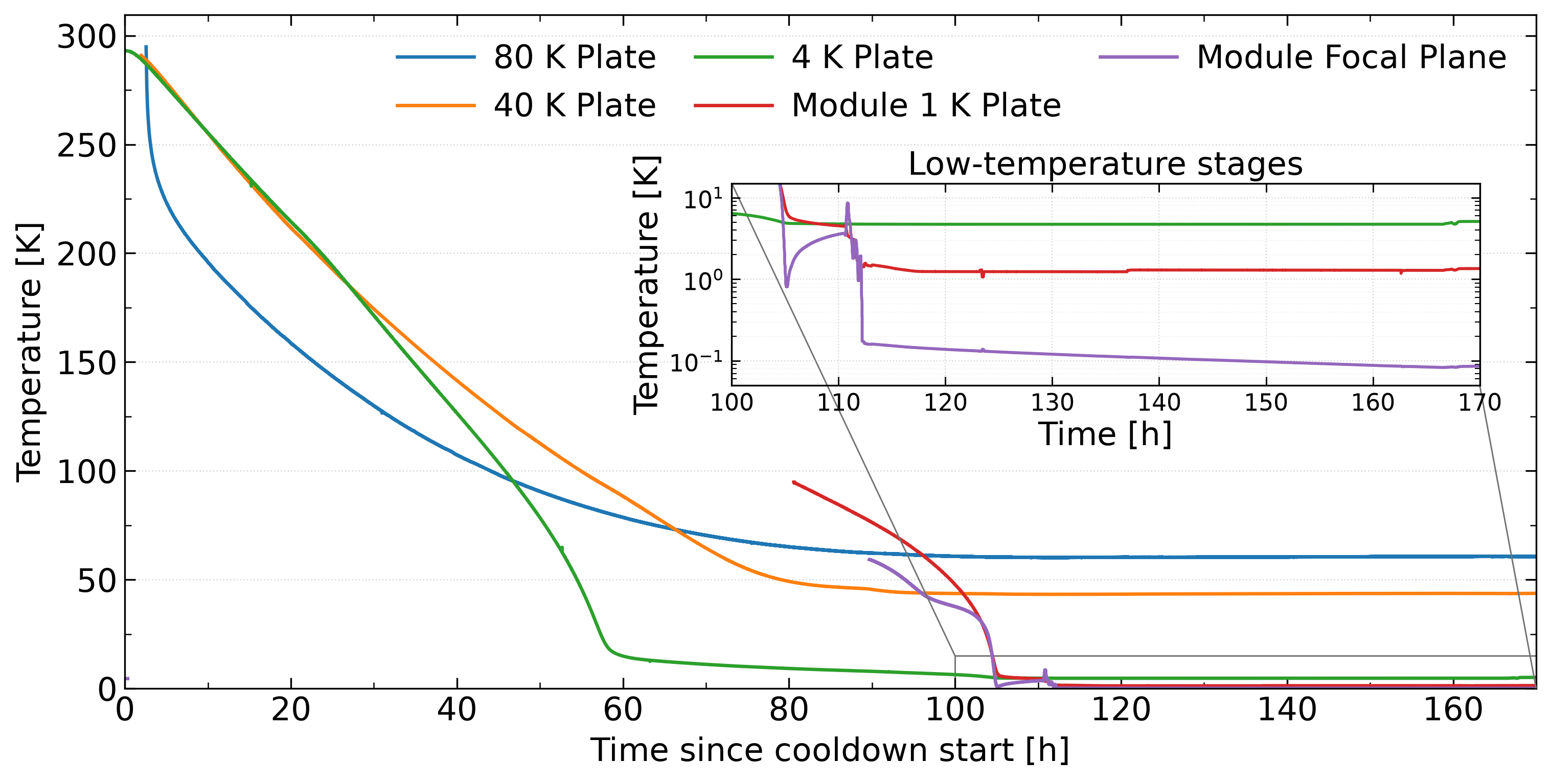}
    \vspace{0.125cm}
    \caption{Representative cooldown curves for the Prime-Cam 80 K, 40 K, 4 K, instrument module 1 K, and instrument module focal plane stages during laboratory testing. The inset shows the low-temperature stages on a logarithmic scale during the mixture condensation and final portion of the cooldown.}
    \vspace{0.125cm}
    \label{fig:cooldown_curve}
\end{figure}

With the fully assembled and populated 280 GHz instrument module installed, and the cryostat optically open to the laboratory environment, Prime-Cam requires slightly less than 5 days ($\sim$110 hours) for its coldest PT-cooled stage to reach 4\,K. Condensation of the DR mixture then requires $\sim$3 hours for the mixing chamber to cool below 100\,mK. However, because the 1 K and 100 mK stages within the instrument module cool relatively slowly, approximately 48 additional hours are required for the system to reach its base temperatures. In total, the full cooldown requires approximately 7 days, or 160 hours, at which point the instrument module focal plane reaches $\sim$80\,mK and the DR mixing chamber reaches $\sim$45\,mK. Turning on all 18 of the 4 K low noise amplifiers increases the focal plane temperature to 88\,mK. Fig.~\ref{fig:cooldown_curve} shows the cooldown curves for the different temperature stages of Prime-Cam. Multiple thermometers were installed at various locations on each stage to monitor both the absolute temperatures and the temperature gradients across the receiver. Here, we present representative measurements from each stage obtained during laboratory testing.

During laboratory testing, heaters were installed on the 80 K, 40 K, 4 K, 1 K, and 100 mK stages to characterize the thermal performance of Prime-Cam. One critical thermal test was performed to estimate the heat load contributed by a single instrument module to the DR mixing chamber. This load was determined by measuring the DR mixing chamber plate temperature under several applied heat loads while maintaining a constant mixture flow rate. The resulting thermal response was then used to project the expected loading from additional instrument modules with thermal characteristics similar to those of the 280 GHz module. We estimate that, in addition to approximately $20\, \mu\mathrm{W}$ of radiative loading from within the enclosed 4 K shell environment, the 280 GHz module contributed approximately $70\,\mu\mathrm{W}$ of heat load to the DR mixing chamber and about 12\,mW to the DR still (1\,K) plate. Under an applied heat load equivalent to that expected from two such modules, the focal plane temperature reached 98\,mK. With an applied heat load equivalent to that expected from four modules, the focal plane temperature reached 110\,mK. Additional details from the laboratory thermal testing campaign will be presented in future publications.

\subsection{Optical Testing}

We present bandpass measurements as a demonstration of the end-to-end optical characterization of Prime-Cam and the 280 GHz module in the laboratory. Optical measurements were performed primarily using a beam mapper and a Fourier transform spectrometer (FTS). The FTS used for these measurements was a Martin–Puplett interferometer in a Mach–Zehnder configuration, based on the PIXIE design \cite{Pan2019}. The instrument and its implementation for receiver testing are described in Ref.~\citenum{patel2025ccatreadout10000280}.

The measurement procedure described in Ref.~\citenum{patel2025ccatreadout10000280} was used here. An approximately 700 K blackbody served as one of the input sources, while a room-temperature blackbody served as the second input source. Measurements were performed on all three 280 GHz detector arrays with the FTS output beam centered at the corresponding position near the cryostat window for each array. The detectors were read out using five RFSoCs through the readout chain shown in Fig.~\ref{fig:280inprimecam}, and the projected detector positions at the cryostat window were determined from the beam-mapping data.

\begin{figure}
    \centering
    \includegraphics[width=0.95\linewidth]{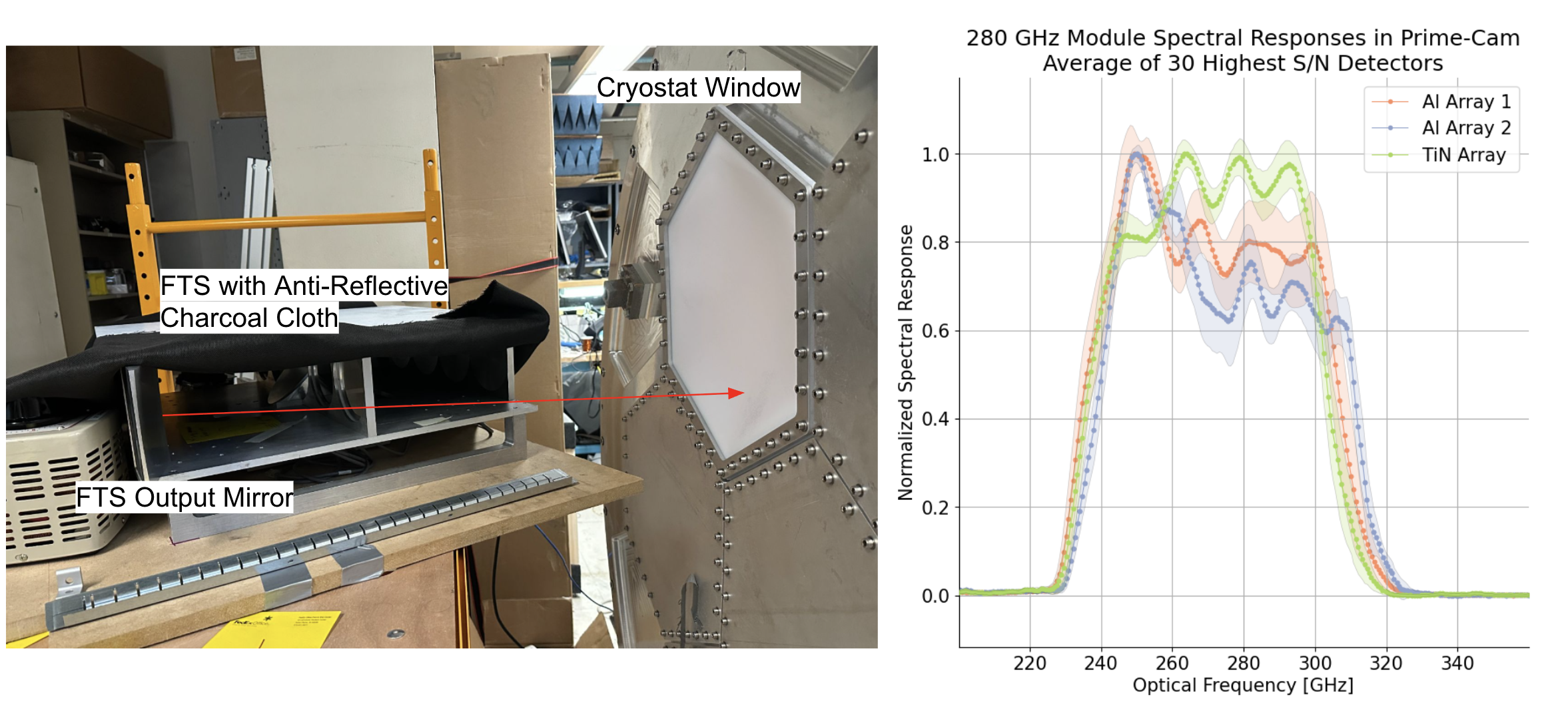}
    \caption{{\it Left}: Fourier transform spectrometer (FTS) setup positioned in front of the optically open Prime-Cam receiver for measurements of the 280 GHz instrument module. Charcoal cloth was placed over the FTS to reduce stray reflections. For each measurement, the FTS output is centered on a small group of detectors and repositioned to probe each of the three detector arrays. The red arrow indicates the optical path from the FTS output mirror to the cryostat window. {\it Right}: Measured spectral response of the 280 GHz instrument module through the complete Prime-Cam optical stack. The bandpasses were measured using KIDs from all three detector arrays.}
    \vspace{0.125cm}
    \label{fig:280fts}
\end{figure}

Figure~\ref{fig:280fts} shows the FTS measurement setup and the measured bandpasses averaged over the 30 detectors with the highest signal-to-noise ratios in each of the three arrays. All three arrays show an approximately 80-GHz-wide bandpass centered near 270\,GHz. We note that the current measurement setup is limited by the alignment of the FTS relative to the cryostat window, including both the input angle and the position of the output beam relative to the selected detectors. Due to these alignment factors, differences in arrays cannot be determined from this data alone. Nevertheless, the laboratory measurements are sufficient for the present purposes of confirming the detector bandpasses to within approximately 5\,GHz and validating the end-to-end optical and readout chains through the successful detection of a well-defined spectral signal. Future measurements would benefit from incorporating dedicated alignment and focusing optics into the FTS setup. We also plan to repeat the bandpass measurements at the CCAT site using a more integrated FTS system.

\section{Early Science with Prime-Cam on FYST}\label{sec:earlysci}

The first year of observations with Prime-Cam on FYST will focus on calibrating the instrument, testing and establishing survey strategies and remote observations, and delivering key science goals for each science team with a subset of Prime-Cam's planned modules. In Sec.~\ref{sec:commissioning} we discuss commissioning and calibration plans for Prime-Cam. In Sec.~\ref{sec:scheduling}, we touch on survey strategy and scheduling for Prime-Cam. Section~\ref{sec:scigoals} details the first year science plans for Prime-Cam with the 280 GHz and 350 GHz modules, and Sec.~\ref{sec:datahandling} summarizes data handling plans.

\subsection{Commissioning and Calibration}
\label{sec:commissioning}

The main aims of the commissioning plan for Prime-Cam are to demonstrate baseline sensitivities for the fielded instrument modules \cite{Choi_2020}, and to establish and test scan strategies for observations. Initial tests upon installing Prime-Cam into FYST's Instrument Space 1 (Fig.~\ref{fig:fyst}) will focus on concurrent testing and observations with a goal of quickly beginning data collection. These tests will include an initial cooldown with the 280 and 350\,GHz modules to check the receiver cryogenics. This will be immediately followed by stationary dark tests to establish detector yields and stable base temperatures. Scanning tests will assess mechanical and cryogenic stabilities across scan speeds and telescope turnarounds to ensure operability. Optical tests will begin with planet observations (Uranus, Neptune, and/or Mars), which will calibrate pointing offsets and stability \cite{Choi_2020}, Prime-Cam focusing and alignment using the translation stage and raft interface (Fig.~\ref{fig:primecamonraft}), detector mapping \cite{Middleton2025}, optimal tone settings \cite{patel2025ccatreadout10000280}, detector responsivity \cite{gazda2025ccatopticalresponsivitynoise}, instrument beams, and end-to-end efficiencies. These calibration observations will be done in parallel with survey scans to prioritize early data taking on FYST.

Environmental conditions on Cerro Chajnantor can affect site work and remote observing plans, in addition to instrument performance when observing. Site conditions, including weather metrics such as air temperature, pressure, humidity, and wind speed, are recorded by a weather station every 15 minutes. Additionally, absorption and thermal emission from water vapor in the atmosphere creates time-varying detector loads and complex added noise. Precipitable water vapor (PWV) in the direction of the science beam will be monitored in real time by an LHATPRO from RPG Radiometer physics GmbH\footnote{\url{https://www.radiometer-physics.de}}, a tropospheric ultra-low humidity and temperature profiler. This radiometer was installed at the site in May 2026, and the first PWV data is currently being studied and compared to other nearby radiometers at lower elevations. 

Baseline module sensitivities will be established from these initial tests and observations, and will address uncertainties in atmospheric loading and instrument effects. At the end of the first observing season, passbands will be measured with an FTS to establish that they meet science-driven requirements\cite{sierra2024simonsobservatorypredeploymentperformance}. Polarization angle uncertainty requirements are similarly determined by science cases, and pointing measurements will be used to obtain optical corrections through polarization-sensitive raytracing with CODE V \cite{Murphy2024,Koopman_2016}.

\subsection{Scheduling and Surveys}\label{sec:scheduling}

Prime-Cam's field of view rotates with both the parallactic angle and the elevation of the telescope (Fig.~\ref{fig:fyst}), and work is ongoing to optimize scan strategies for the instrument \cite{ccat2023}. Constant elevation scans, during which the telescope scans in azimuth at constant elevations, then turns around and scans in the opposite direction, have been well-established by the Atacama Cosmology Telescope (ACT) and SO \cite{Stevens_2018}. The Prime-Cam wide field scan strategy will employ a similar strategy. Improvements to the typical constant elevation scans include modulation of azimuthal angular velocity and sinusoidal elevation nods, which improve survey uniformity and crosslinking while enabling higher cadence observations for time-domain astrophysics \cite{Ebina_2022}. 

Other telescope scanning patterns are also being studied, with mapping parameters tuned based on telescope kinematics analysis,
scan homogeneity inside the area of interest, and mapping efficiency. Understanding the observable elevation range accessible without re-tuning the detectors will motivate future investigation and scan pattern adaptations. Prime-Cam surveys will be conducted by remote observers using a scheduler, designed to be more flexible than state-of-the-art static-CSV schedulers \cite{guan2024simonsobservatoryobservatoryscheduler}. The scheduler combines: (1) calibration and science survey requirements and priorities; (2) instrument status information; (3) available targets and sky regions for observations at a given time including Sun and Moon avoidance; and (4) the current (and potentially predicted) PWV to dynamically suggest the next target to observe. We use the same algorithm to simulate various weather scenarios over a multi-week scheduling block using historical PWV templates and optimize tunable parameters in the scheduling algorithm (Okada et al. in prep.).

\begin{table*}
\centering
\renewcommand{\arraystretch}{1.5}
\caption{Prime-Cam science cases, survey names, and planned early science survey time as compared to the full 5-year Prime-Cam survey \cite{ccat2023}.}
\vspace{0.125cm}
\begin{tabular}{
>{\raggedright\arraybackslash}p{2.15cm}
>{\raggedright\arraybackslash}p{6.4cm}
c c c
}
\toprule
Survey &
Science case &
\makecell[c]{Survey Time,\\ Early Science [hr]} &
\makecell[c]{Survey Time,\\ 5-Year Survey [hr]}\\
\midrule

\dots &
Calibration (planet observations) &
2 obs/week &
TBD\\

WFS &
\makecell[tl]{CMB, SZ, galaxy evolution,\\
Rayleigh scattering, time domain} &
200 &
2,000\\

CIB &
Galaxy evolution &
100 &
500\\

GalPol &
\makecell[tl]{Galactic dust polarization: magnetic \\
fields, star formation, dust models} &
160 &
525 \\

Time domain &
\makecell[tl]{Variable Galactic protostars;\\
Galactic center monitoring;\\
transients} &
\makecell[tc]{0.3/epoch;\\
Target-of-Opp.} &
\makecell[tc]{500;\\
Target-of-Opp.} \\

\bottomrule
\end{tabular}
\label{tab:modulesurveys}
\end{table*}

\subsection{Science Goals}\label{sec:scigoals}

The primary goals of the early science observation period are to deliver new results for each Prime-Cam science case (Table~\ref{tab:modulesurveys}), while preparing for the full nominal 5-year Prime-Cam survey \cite{ccat2023}. 

\subsubsection{Cosmic Microwave Background (CMB)}

Observations for the CMB foregrounds, Rayleigh scattering, and Sunyaev-Zeldovich science cases will use the Wide Field Survey (WFS) to: (1) Reduce bias on tensor-to-scalar ratio constraints from CMB surveys to aid in the search for primordial gravitational waves; (2) study galaxy group and cluster evolution through the Sunyaev-Zeldovich Effects; (3) study the polarization of the Milky Way to better understand star-formation processes; (4) search for Rayleigh scattering of the CMB, a small alteration of the CMB blackbody spectrum embedding information about the universe's expansion history; and (5) produce blind detections and follow-up measurements of transient events \cite{ccat2023}. Early science data will be collected over approximately 200 hours of nighttime observations in a small region ($\sim$400 sq. deg.) overlapping with the ACT ``Deep56'' region \cite{De_Bernardis_for_the_ACT_collaboration_2016} at full WFS survey depth. Repeated observations of this smaller region while varying scan strategies will aid in understanding systematics while achieving targeted map depth. This region will overlap completely with Simons Observatory and ACT data and other surveys such as the Baryon Oscillation Spectroscopic Survey (BOSS) \cite{Dawson_2012}, Hyper Suprime-Cam (HSC) \cite{Aihara_2017}, the Dark Energy Survey \cite{thedarkenergysurveycollaboration2005darkenergysurvey}, and the Dark Energy Spectroscopic Instrument, \cite{desicollaboration2016desiexperimentisciencetargeting} enabling rich cross-correlation opportunities.

\subsubsection{CIB Galaxy Evolution in Extragalactic Fields}

The early science program for the cosmic infrared background (CIB) galaxy evolution science case will survey up to 1000 sq. deg. across ten priority-ranked fields, in $\sim$100 hours of observing time. Field priorities will be optimized for prevailing weather conditions. Mapping these fields will verify that the survey reaches the extragalactic confusion limit at the instrument resolution, recover large-scale galaxy clustering on angular scales exceeding $1^\circ$, and establish accurate point-source flux density calibration through comparison with existing surveys at similar wavelengths. The selected fields overlap extensively with {\it Herschel}-SPIRE \cite{Griffin_2010} and VLA/MeerKAT \cite{deka2023meerkatabsorptionlinesurvey} observations, enabling multiwavelength characterization of dusty star-forming galaxies and active galactic nuclei. These early science observations will also probe previously unexplored regions of survey parameter space by combining wide sky coverage, high sensitivity, and large angular scales, enabling new measurements of the CIB, the clustering of dusty star-forming galaxies, and the connection between obscured star formation and large-scale structure \cite{ccat2023}.

\subsubsection{Galactic Polarization}

The early science Galactic Polarization program will study star formation, magnetic fields, turbulence, and dust grain properties in the Milky Way and nearby galaxies \cite{ccat2023,Hensley2022}. The program will target up to four of a rank-ordered list of molecular cloud regions, with each field requiring approximately 30--50 hours of observing time. Observations are designed to demonstrate Prime-Cam's polarimetric performance while producing scientifically valuable maps of magnetic field structure in different types of Milky Way star-formation environments \cite{ccat2023}. Each field will be observed over a range of elevations and parallactic angles to maximize the diversity of directly measured polarization orientations, enabling robust structure characterization. Comparisons between daytime and nighttime observations will quantify the impact of scheduling on polarization data quality. The baseline scan strategy is constant-elevation scans, while a ``curvy pong'' scanning strategy \cite{Holland2013} is undergoing feasibility studies to provide uniform sky coverage and cross-linked scans for accurate recovery of extended polarized emission in case detector stability will allow for the strategy \cite{Ebina_2022}. The selected fields overlap with existing SOFIA/HAWC+ \cite{HAWC2018}, BLASTPol \cite{Galitzki_2014}, James Clerk Maxwell Telescope (JCMT) POL-2 \cite{Pattle2017}, and {\it Planck} \cite{Planck2015} polarization observations, enabling cross-validation of polarization measurements and inferred magnetic field properties. The  calibration program for Galactic polarization will include planet observations for beam characterization, measurements of unpolarized sources such as planets and asteroids to quantify instrument polarization, and observations of bright polarized calibration quasars from the ALMA polarization calibrator catalog to determine the instrument's polarization efficiency and polarization angle.

\subsubsection{Time-domain Science}

Regularly repeated wide-field observations at submillimeter wavelengths will enable Prime-Cam to investigate many types of variable astronomical source, including asteroids, supernovae, gamma-ray bursts, and tidal disruption events.  In particular, the time-domain science program will offer new insights into the variability in protostars within Galactic star-forming regions and monitoring of the Galactic center, while establishing transient follow-up observations \cite{ccat2023}. A prioritized target list of Galactic star-forming regions has been compiled, with each field requiring around 20 minutes of observing time per epoch and revisited on a monthly cadence. The two-week cadence, repeated observations will measure continuum variability associated with episodic accretion and other time-dependent processes in young stellar objects and the Galactic center. Consistent with established methods at the JCMT \cite{jcmttransients}, the fields will be self-calibrated to achieve the expected $1\%$ relative photometric precision required for Prime-Cam's long-term variability studies. In addition to the scheduled monitoring program, the observing plan includes target-of-opportunity observations of transient events discovered by the Simons Observatory \cite{ASO2025} or other facilities. These triggered observations will require 1 hour per epoch, and will interrupt scheduled observations to provide rapid submillimeter follow-up of newly identified transients. Furthermore, the WFS observations will be investigated to search for new transient sources.

\subsection{Software and data handling}\label{sec:datahandling}

A largely integrated framework for remote observing has been developed for Prime-Cam that builds on the SO Observatory Control Software (OCS) infrastructure \cite{koopman2024simonsobservatorydeploymentobservatory,Nguyen_2024} while incorporating Prime-Cam-specific interfaces and workflows. This ``Prime-Cam Control System" (PCS)\footnote{\url{https://github.com/ccatobs/pcs};   \url{https://grafana.com}} combines the telescope scheduler, a custom operations database (OpsDB), and the Nextline web interface to automatically generate executable observing scripts from scheduled observations, while allowing for observer intervention and script editing when needed \cite{middleton2026}. Detector tuning, calibration, telescope control, and data acquisition are orchestrated within a single observing script, and the system has been demonstrated with simulated telescope scans and RFSoC detector readout in the lab.

Remote Prime-Cam operations are supported by comprehensive monitoring and data management tools. Instrument housekeeping, receiver and telescope status are continuously logged and displayed through Grafana$^{\dagger\dagger}$. dashboards with automated Slack alerts for anomalous conditions, while the PCS-web interface provides browser-based remote control of hardware, instrument processes, and telescope functions without requiring direct interaction with low-level software. Observing data are automatically registered with OpsDB, transferred to on-site analysis computers for quick-look data quality assessment, and archived for off-site processing and long-term storage at the CCAT Data Center at the University of Cologne \footnote{\url{docs.data.ccat.uni-koeln.de}}. Scalable data-reduction pipelines are being developed and tested on the high-performance computing (HPC) systems at the CCAT Data Center to process raw detector timestreams into calibrated, science-ready data products.

\section{Conclusion}

Prime-Cam on the CCAT Observatory's Fred Young Submillimeter Telescope will enable transformative wide-field and targeted observations of the submillimeter sky from the exceptional 5600 m Cerro Chajnantor site. The receiver has been successfully laboratory tested, along with its first-light 280 GHz and 350 GHz broadband polarimetric instrument modules, validating the cryogenic, mechanical, optical, detector, and readout systems in advance of deployment and telescope integration. In parallel, commissioning procedures, calibration strategies, remote observing infrastructure, and early science survey plans have been developed with the goal of rapidly transitioning from telescope integration to scientific observations.

Prime-Cam has now been shipped to Chile for installation on FYST, with integration and commissioning scheduled for 2026, followed by the start of early science observations. The first year of observations will establish the instrument's on-sky performance while delivering new measurements spanning cosmology, galaxy and structure evolution, Galactic magnetic fields, and time-domain astrophysics. As additional instrument modules are commissioned over the coming years, Prime-Cam will field over 100,000 kinetic inductance detectors across the millimeter and submillimeter bands, opening new windows on the evolution of the Universe.

\acknowledgments 
The CCAT project, FYST and Prime-Cam instrument have been supported by generous contributions from the Fred M. Young, Jr. Charitable Trust, Cornell University, Duke University, the Canada Foundation for Innovation, and the Provinces of Ontario, Alberta, and British Columbia. The construction of the FYST telescope was supported by the Großgeräte-Programm of the German Science Foundation (Deutsche Forschungsgemeinschaft, DFG) under grant INST 216/733-1 FUGG, as well as funding from Universität zu Köln, Universität Bonn, and the Max Planck Institut für Astrophysik, Garching. The construction of EoR-Spec is supported by NSF grant AST-2009767. The construction of the 350 GHz instrument module for Prime-Cam is supported by NSF grant AST-2117631. The construction of the 850 GHz instrument module for Prime-Cam is supported by CFI grants $\#$39656 and $\#$46097 and Canadian provincial matching funds. The completion and deployment of the Prime-Cam instrument with the initial instrument modules is supported by a generous contribution from Alex Gerko, Founder and CEO of XTX Markets. This work has been supported by the Collaborative Research Center 1601 (SFB 1601 sub-project C3) funded by the Deutsche Forschungsgemeinschaft (DFG, German Research Foundation) -- 500700252. MA is supported by FONDECYT grant 1252054, and gratefully acknowledges support from ANID Basal Project FB210003,  ANID MILENIO NCN2024112 and ANID + Vinculacion Internacional + FOVI250261. D.R. gratefully acknowledges support from the
Collaborative Research Center 1601 (SFB 1601 sub-projects C1, C2, C3,
and C6) and through the Cluster of Excellence ``Our Dynamic Universe''
under Germany's Excellence Strategy, both funded by the Deutsche
Forschungsgemeinschaft (DFG) – 500700252, and EXC 3037 – 533607693. D.J.\ is supported by NRC Canada and by an NSERC Discovery Grant. RGF acknowledges support from NSF ATI Grant 2009767. GJS acknowledges support from NSF ATI Grant 2009767. TN acknowledges support from NSF ATI Grant 2009767. ATC acknowledges support as a Fred Young Faculty Fellow.

% References
\bibliography{nab_aag}

\bibliographystyle{spiebib} % makes bibtex use spiebib.bst

\end{document}